\documentclass[a4paper,oneside,11pt]{article}
\usepackage{amsmath}
\usepackage{amssymb}
\usepackage{amsthm}
\usepackage{enumerate}
\usepackage{bm}
\usepackage{graphicx}
\usepackage{makecell} 
\usepackage{enumitem} 
\usepackage[margin = 3cm]{geometry}
\usepackage{hyperref,url}

\renewcommand{\vec}{\bm}

\newcommand{\Div}{{\rm div}}
\newcommand{\abs}[1]{\left| #1 \right|}
\newcommand{\eps}{\varepsilon}

\newtheorem{thm}{Theorem}[section]

\newtheorem{defn}{Definition}[section]
\newtheorem{assump}{Assumption}[section]

\allowdisplaybreaks

\begin{document}

\title{Two-phase flow with surfactants: Diffuse interface models and their analysis}

\date{}

\author{Helmut Abels \footnotemark[1] \and Harald Garcke \footnotemark[1] \and Kei Fong Lam \footnotemark[1] \and Josef Weber \footnotemark[1]}

\renewcommand{\thefootnote}{\fnsymbol{footnote}}
\footnotetext[1]{Fakult\"at f\"ur Mathematik, Universit\"at Regensburg, 93040 Regensburg, Germany
({\tt \{Helmut.Abels, Harald.Garcke, Kei-Fong.Lam, Josef.Weber\}@mathematik.uni-regensburg.de}).}

\maketitle

\begin{abstract}
New diffuse interface and sharp interface models for soluble and insoluble surfactants fulfilling energy inequalities are introduced.  We discuss their relation with the help of asymptotic analysis and present an existence result for a particular diffuse interface model.
\end{abstract}

\section{Introduction}\label{sec:Intro}

Surface active agents (or commonly known as surfactants) are compounds that are able to lower the surface tension between fluidic interfaces, and thus have found numerous applications in both biological systems and industrial processes. 

For systems with two or more immiscible fluids, surfactants can be broadly classified into two types: insoluble and soluble.  In the latter case, surfactants can exist in both the bulk fluid phases and also on the fluid interfaces, but in the former case, insoluble surfactants will only exist on the interfaces.  When introduced to a multi-fluid system, the soluble surfactants may migrate towards the fluid interfaces and are incorporated to the interface by the process of adsorption.  

One of the simplest model of adsorption dynamics is that studied by Ward and Tordai \cite{WardTordai} and is defined on $(0,\infty)$ with the interface at the origin:
\begin{align*}
\partial_{t}c = D \partial_{xx} c \; \mbox{ for }  \; x > 0, \; t > 0, \quad \partial_{t}c^{\Gamma} = D \partial_{x} c \; \mbox{ at } \; x = 0, \; t > 0, \\
\lim_{x \to \infty} c(x,t) = c_{b} \; \mbox{ for } \; t > 0, \quad c(x,0) = c_{b}, \quad c^{\Gamma}(0) = 0.
\end{align*}
Here, $c$ and $c^{\Gamma}$ denote the concentration of the bulk and interfacial surfactants, $c_{b}$ is the initial and far-field boundary condition, $D$ denotes the diffusion coefficient, and the term $D \partial_{x} c$ is a source term for the surfactant concentration on the interface stemming from the surfactant flux in the bulk.  In their work, Ward and Tordai derived an equation relating the interfacial surfactant concentration $c^{\Gamma}(t)$ and the surfactant concentration of the sub-layer $c(0,t)$, where the sub-layer is defined as the bulk region immediately adjacent to the interface.  To solve for $c^{\Gamma}(t)$, Ward and Tordai assumed that the sub-layer and the interface are in thermodynamical equilibrium, and thus postulates a relation between $c^{\Gamma}(t)$ and $c(0,t)$, which is given by
\begin{equation}\label{WT:isotherm}
c^{\Gamma}(t) = g(c(0,t))
\end{equation}
for some function $g$.  This functional relation is termed \emph{adsorption isotherm} \cite{EastoeDalton}, which relates the interfacial concentration with the sub-layer concentration.

A key assumption in the work of Ward and Tordai is that the interface and the sub-layer are in equilibrium, that is, the process of adsorption is fast compared to the kinetics in the bulk regions.  This case is called \emph{instantaneous adsorption} or \emph{diffusion controlled adsorption}.  However, there are systems in which instantaneous adsorption is not valid, for example in the context of ionic surfactants \cite{DiamantAndelman}, and in these situations a closure relation akin to (\ref{WT:isotherm}) between $c(0,t)$ and $c^{\Gamma}(t)$ is not available.  For such cases, which we denote as \emph{non-instantaneous adsorption} or \emph{dynamic adsorption}, we will have to postulate alternative closure relations.

Two-phase flow with surfactant is classically modelled with moving hypersurfaces describing the interfaces separating the two fluids.  In \cite{GarckeLamStinner}, the following sharp interface model for a domain $\Omega$ containing two fluids of different mass densities in the presence of soluble surfactants is derived.  We denote by $\Omega_{-}(t)$, $\Omega_{+}(t)$ the domains of the fluids which are separated by an interface $\Gamma(t)$: 
\begin{subequations}\label{GLS:SIM}
\begin{alignat}{3}
\Div \, \vec{v} & =  0 && \text{ in } \Omega_{\pm}(t), \label{SIM:incompress} \\
\partial_{t}(\tilde{\rho}_{\pm} \vec{v}) + \Div \, (\tilde{\rho}_{\pm} \vec{v} \otimes \vec{v}) & =  \Div \, \left ( - p Id + 2\eta_{\pm}  \vec{D} \vec{v} \right )&& \text{ in } \Omega_{\pm}(t), \label{SIM:momentum}\\
\partial_{t}^{\bullet} c_{\pm} & =  \Div \, (M_{c}^{\pm} \nabla G_{\pm}'(c_{\pm})) &&  \text{ in } \Omega_{\pm}(t), \label{SIM:bulk} \\
[\vec{v}]_{-}^{+} & =  0, \quad \vec{v} \cdot \vec{\nu} = \mathcal{V} && \text{ on } \Gamma(t), \label{SIM:velocityjump}\\
[p Id - 2\eta \vec{D} \vec{v}]_{-}^{+} \vec{\nu} & =  \sigma(c^{\Gamma}) \kappa \vec{\nu} + \nabla_{\Gamma} \sigma(c^{\Gamma}) && \text{ on } \Gamma(t), \label{SIM:stressjump} \\
\partial_{t}^{\bullet} c^{\Gamma} + c^{\Gamma} \, \Div_{\Gamma} \, \vec{v} - \Div_{\Gamma} \, (M_{\Gamma} \nabla_{\Gamma} \gamma'(c^{\Gamma})) & = [M_{c} \nabla G'(c)]_{-}^{+} \vec{\nu} && \text{ on } \Gamma(t), \label{SIM:interface} \\ 
\mp \alpha_{\pm}  M_{c}^{\pm} \nabla G'_{\pm}(c_{\pm}) \cdot \vec{\nu} & = -(\gamma'(c^{\Gamma}) - G_{\pm}'(c_{\pm})) && \text{ on } \Gamma(t). \label{SIM:dynamicAdsorp}
\end{alignat}
\end{subequations}

Here $\vec{v}$ denotes the fluid velocity, $\tilde{\rho}_{\pm}$ and $\eta_{\pm}$ are the constant mass densities and viscosities of the fluids, respectively, $\vec{D} \vec{v} = \tfrac{1}{2}(\nabla \vec{v} + (\nabla \vec{v})^{\top})$ is the rate of deformation tensor, $p$ is the pressure, $Id$ is the identity tensor, $\partial_{t}^{\bullet}(\cdot) = \partial_{t}(\cdot) + \vec{v} \cdot \nabla (\cdot)$ is the material derivative, $c_{\pm}$ are the bulk densities of the surfactants, $M_{c}^{\pm}$ are the bulk mobilities, and $G_{\pm}$ are the bulk free energy densities.  

On the interface, $\mathcal{V}$ is the normal velocity, $\vec{\nu}$ is the unit normal on $\Gamma$ pointing into $\Omega_{+}$, $c^{\Gamma}$ is the interfacial surfactant density, $\gamma(c^{\Gamma})$ is the interfacial free energy density, $\sigma(c^{\Gamma}) := \gamma(c^{\Gamma}) - c^{\Gamma} \gamma'(c^{\Gamma})$ is the density dependent surface tension, $\kappa$ is the mean curvature of $\Gamma$, $\nabla_{\Gamma}$ is the surface gradient operator, $\Div_{\Gamma}$ is the surface divergence, $M_{\Gamma}$ is the interfacial mobility, and $\alpha_{\pm} \geq 0$ are kinetic factors which are related to the speed of adsorption. 

Equations \eqref{SIM:incompress} and \eqref{SIM:momentum} are the classical incompressibility condition and momentum equation, respectively.  The mass balance equation for bulk surfactants is given by \eqref{SIM:bulk}.  Equation \eqref{SIM:velocityjump} states that the interface is transported with the flow and that not only the normal components but also the tangential components of the velocity field match up. The force balance on the interface \eqref{SIM:stressjump} relates the jump in the stress tensor across the interface to the surface tension force and the Marangoni force at the interface.  The mass balance of the interfacial surfactants is given by \eqref{SIM:interface}, and the closure condition \eqref{SIM:dynamicAdsorp} tells us whether adsorption is instantaneous ($\alpha = 0$, an isotherm is obtained) or dynamic ($\alpha > 0$, the mass flux into the interface is proportional to the difference in chemical potentials).  

To see this, suppose the process of adsorption at the interface is instantaneous, i.e., fast compared to the timescale of convective and diffusive transport in the bulk regions.  This local equilibrium corresponds to the case that the bulk chemical potential $G_{\pm}'(c)$ and the interface chemical potential $\gamma'(c^{\Gamma})$ are equal, which is the case if we set $\alpha = 0$ in \eqref{SIM:dynamicAdsorp} (we here only consider one of the bulk phases adjacent to the interface and, for simplicity, drop the subscript $\pm$). We obtain the following relation 
\begin{align}\label{eq:instadsorption}
\gamma'(c^{\Gamma})  = G'(c) \quad \Longleftrightarrow \quad  c^{\Gamma} = g(c) := (\gamma')^{-1}(G'(c)),
\end{align}
where $g: \mathbb{R}_{+} \to \mathbb{R}_{+}$ is strictly increasing.  This function $g$ plays the role of various adsorption isotherms which state the equilibrium relations between the two densities.  The novelty of the model \eqref{GLS:SIM} is two-fold; we can realise various adsorption isotherms by choosing appropriate functional forms for the free energy densities $G$ and $\gamma$, see Table \ref{tbl:Isotherms} below and also \cite[Table 2.1]{GarckeLamStinner} for examples.  Moreover, with positive values of $\alpha$, we can include the effects of non-equilibrium adsorption dynamics, and thus \eqref{GLS:SIM} is a generalisation of the model studied in \cite{BothePruss,BothePrussSimonett} to the case of dynamic adsorption.  Furthermore a model involving insoluble surfactants easily arises by setting $c_{\pm} = M_{c}^{\pm} = 0$ and neglecting \eqref{SIM:dynamicAdsorp} in \eqref{GLS:SIM}.

In Table \ref{tbl:Isotherms} the functional forms for $\gamma$ and $G$ for the Henry and Langmuir adsorption isotherms are stated. Here, $c^{\Gamma}_{M}$ is the maximum interfacial surfactant density, $K$ is a constant relating the surface density to the bulk density in equilibrium, $\sigma_{0}$ denotes the surface tension of a clean interface, and $B$ is the sensitivity of the surface tension to surfactant.  We point out that the model \eqref{GLS:SIM} satisfies the second law of thermodynamics in an isothermal situation in the form of an energy dissipation inequality (under suitable boundary conditions), cf. \cite{GarckeLamStinner,GarckeWieland},
\begin{equation}\label{SIM:Energy:Ineq}
\begin{aligned}
0 & = \frac{d}{dt} \left ( \left [\sum \int_{\Omega_{\pm}}  \left ( \tfrac{\tilde{\rho}_{\pm}}{2} \abs{\vec{v}}^{2} + G_{\pm}(c_{\pm})\right )  \right ] +  \int_{\Gamma} \gamma(c^{\Gamma}) \right ) + \int_{\Gamma}  M_{\Gamma} \abs{\nabla_{\Gamma} \gamma'(c^{\Gamma})}^{2}  \\
& + \sum \left ( \int_{\Omega_{\pm}} \left ( \eta_{\pm} \abs{\vec{D} \vec{v}}^{2} + M_{c}^{\pm} \abs{\nabla G_{\pm}'(c_{\pm})}^{2} \right ) + \int_{\Gamma} \frac{1}{\alpha_{\pm}} \abs{\gamma'(c^{\Gamma}) - G_{\pm}'(c_{\pm})}^{2} \right ).
\end{aligned}
\end{equation}
The model \eqref{GLS:SIM} constitutes a free boundary problem, in which the interface $\Gamma(t)$ is unknown a priori and has to be computed as part of the solution.  For numerical simulations of two-phase flow with surfactants based on the above models, we refer the reader to the work of \cite{BGN:Insoluble,BGN:Soluble}.

\begin{table}[t]
\centering
\begin{tabular}{|c|c|c|}
\hline
Isotherm & Henry & Langmuir \\
\hline 
Relation  & $Kc = \frac{c^{\Gamma}}{c^{\Gamma}_{M}}$ & \gape{$Kc = \frac{c^{\Gamma}}{c^{\Gamma}_{M} - c^{\Gamma}}$} \\ [2ex]
\hline
$\gamma(c^{\Gamma}) - \sigma_{0}$ & $ B c^{\Gamma}(\log \frac{c^{\Gamma}}{c^{\Gamma}_{M}} -1)$ & \gape{$B \left (c^{\Gamma} \log \frac{c^{\Gamma}}{c^{\Gamma}_{M} - c^{\Gamma}} + c^{\Gamma}_{M} \log ( 1 - \frac{c^{\Gamma}}{c^{\Gamma}_{M}} ) \right ) $} \\ [2ex]
\hline 
$G(c) $ & $Bc(\log(Kc)-1)$ & \gape{$B c (\log(Kc)-1)$}  \\ 
\hline
$\sigma - \sigma_{0} $ & $-B c^{\Gamma}$  & \gape{$Bc^{\Gamma}_{M} \log \left ( 1 - \frac{c^{\Gamma}}{c^{\Gamma}_{M}} \right )$} \\ [2ex]
\hline
\end{tabular}
\caption{Possible functional forms for $\gamma$ and $G$ to obtain the Henry and Langmuir adsorption isotherms and equations of state.  See \cite{GarckeLamStinner} for the details regarding the Freundlich, Volmer and Frumkin isotherms.}
\label{tbl:Isotherms}
\end{table}

A second approach is to model the dynamics is to relax the immiscibility assumption of the fluids, and assume that there are some microscopic mixing of the macroscopically immiscible fluids.  This replaces the hypersurface description with a interfacial layer of small and finite width.  Within this layer the fluids are assumed to be mixed and thus we have to account for the mixing energies.  These models are commonly termed as \emph{diffuse interface} or \emph{phase field models}, and at the core of these models is an order parameter which takes distinct constant values in the bulk phases and varies smoothly across the interfacial layer.  A first diffuse interface model for two-phase flows with matched densities was proposed by Hohenberg and Halperin \cite{HH}.  For different densities, Lowengrub and Truskinowsky \cite{Lowengrub} proposed a thermodynamically consistent quasi-incompressible model based on a mass-averaged velocity which is not solenoidal.  Meanwhile, Abels, Garcke and Gr\"{u}n \cite{AbelsGarckeGrun,AGGCompanion} derived a diffuse interface model for unmatched densities with a solenoidal velocity field based on a volume-averaged velocity.  In contrast to the model of Ding, Spelt and Shu \cite{Ding} which also employs a volume-averaged solenodial velocity field, the model of Abels, Garcke and Gr\"{u}n is thermodynamically consistent and fulfills local and global free energy inequalities.  The goal is to derive thermodynamically consistent diffuse interface models for two-phase flow with soluble surfactants, using the model of Abels, Garcke and Gr\"{u}n \cite{AbelsGarckeGrun} as our basis. 


\section{Diffuse interface models}\label{sec:DIM}
At the core of any diffuse interface model lies the Ginzburg--Landau functional
\begin{align*}
\mathcal{E}(\varphi) := \int_{\Omega} \left ( \frac{\eps}{2} \abs{\nabla \varphi}^{2} + \frac{1}{\eps} \psi(\varphi) \right ).
\end{align*}
Here, $\varphi : \Omega \to \mathbb{R}$ denotes the order parameter used to distinguish the bulk fluid phases, $\eps > 0$ is a parameter related to the thickness of the interfacial layer, and $\psi$ is a potential with two equal minima (which we will take to be $\pm 1$).  Through the work of Modica and Mortola \cite{Modica}, it is well-known that the Ginzburg--Landau functional $\mathcal{E}(\varphi)$ converges to a multiple of the perimeter functional of the set $\{\varphi = 1\}$ in the sense of $\Gamma$-convergence.  Hence $\mathcal{E}$ is often used to approximate the surface energy on the interface.  Let us denote by $\delta_{\Gamma}$ the Hausdorff measure restricted to $\Gamma$, and by $\chi_{\Omega_{\pm}}$ the characteristic function of the set $\Omega_{\pm}$.  With the help of \cite[\S 2.7 and Theorem 2.8]{Alt} (see also \cite[Appendix B]{LamThesis}) the surfactant subsystem \eqref{SIM:bulk}, \eqref{SIM:interface}, \eqref{SIM:dynamicAdsorp} can be reformulated into an equivalent distributional form
\begin{subequations}
\begin{align}
\partial_{t}(\chi_{\Omega_{\pm}} c_{\pm}) + \Div \, (\chi_{\Omega_{\pm}} c_{\pm} \vec{v} - \chi_{\Omega_{\pm}} M_{c}^{\pm} \nabla G_{\pm}'(c_{\pm})) & = \delta_{\Gamma} j_{\pm}, \\
\partial_{t}(\delta_{\Gamma} c^{\Gamma}) + \Div \, (\delta_{\Gamma} c^{\Gamma} \vec{v} -  \delta_{\Gamma} M_{\Gamma} \nabla \gamma'(c^{\Gamma})) & = - \delta_{\Gamma} (j_{-} + j_{+}), \\
j_{\pm} & = \tfrac{1}{\alpha_{\pm}} (\gamma'(c^{\Gamma}) - G_{\pm}'(c_{\pm})) .
\end{align}
\end{subequations}
The idea of \cite{GarckeLamStinner} is to replace the distributions $\delta_{\Gamma}$ and $\chi_{\Omega_{\pm}}$ with regularisations $\delta_{\eps}(\varphi, \nabla \varphi)$ and $\xi_{\pm,\eps}(\varphi)$ indexed by the width of the interfacial layer $\eps > 0$.  This is done in the spirit of the so-called diffuse domain approach \cite{LLRV,Teigen}.  For a rigorous  treatment of the diffuse domain approach in the limit $\eps \to 0$ we refer the reader to \cite{AbelsLamStinner,Burger,Franz}.  One example of a regularisation $\delta_{\eps}$ is the Ginzburg--Landau density
\begin{align*}
\delta_{\eps}(\varphi, \nabla \varphi) = \mathcal{W} \left ( \frac{\eps}{2} \abs{\nabla \varphi}^{2} + \frac{1}{\eps} \psi(\varphi) \right ), \quad \frac{1}{\mathcal{W}} := \int_{-1}^{1} \sqrt{2 \psi(s)} ds.
\end{align*}
In the following, we will rescale the potential $\psi$ so that $\mathcal{W} = 1$.  Meanwhile, we can take $\xi_{-,\eps}(\varphi) = \frac{1-\varphi}{2}$ and $\xi_{+,\eps}(\varphi) = 1 - \xi_{-,\eps}$.  Then, for $\alpha_{\pm} > 0$, the diffuse interface model for soluble surfactants of \cite{GarckeLamStinner} (denoted as Model A) in the case of dynamic adsorption is given as (dropping the subscript $\eps$ from $\xi_{\pm,\eps}$ and $\delta_{\eps}$)
\begin{subequations}\label{GLS:ModelA}
\begin{align}
\Div \, \vec{v} & = 0, \label{PFMA:incompress} \\
\partial_{t}(\rho \vec{v})  + \Div \, ( \rho \vec{v} \otimes \vec{v} ) & = \Div \, \left ( - p Id + 2 \eta(\varphi) \vec{D} \vec{v} + \vec{v} \otimes \tfrac{\tilde{\rho}_{+} - \tilde{\rho}_{-}}{2} m(\varphi) \nabla \mu \right ) \label{PFMA:momentum} \\
\nonumber & \quad + \Div \, \left ( \sigma(c^{\Gamma}) (\delta(\varphi, \nabla \varphi) Id  - \eps \nabla \varphi \otimes \nabla \varphi) \right ), \\
\partial_{t}^{\bullet} \varphi & = \Div \, (m(\varphi) \nabla \mu),  \label{PFMA:phase} \\
\mu + \Div \, ( \eps \sigma(c^{\Gamma}) \nabla \varphi) & = \frac{\sigma(c^{\Gamma})}{\eps} \psi'(\varphi) + \sum \xi'_{\pm}(\varphi)(G_{\pm}(c_{\pm}) - G_{\pm}'(c_{\pm}) c_{\pm}), \label{PFMA:chem}  \\ 
\partial_{t}^{\bullet} (\xi_{\pm}(\varphi) c_{\pm}) & = \Div \, (M_{c}^{\pm}(c_{\pm}) \xi_{i}(\varphi)\nabla G_{i}'(c_{\pm})) \label{PFMA:bulk} \\
\nonumber & \quad + \tfrac{1}{\alpha_{\pm}} \delta(\varphi, \nabla \varphi)(\gamma'(c^{\Gamma}) - G'_{\pm}(c_{\pm})),  \\
\partial_{t}^{\bullet} (\delta(\varphi, \nabla \varphi) c^{\Gamma}) & = \Div \, \left ( M_{\Gamma}(c^{\Gamma}) \delta(\varphi, \nabla \varphi) \nabla \gamma'(c^{\Gamma}) \right )  \label{PFMA:interface}  \\
\nonumber & \quad - \delta(\varphi, \nabla \varphi) \sum \tfrac{1}{\alpha_{\pm}}(\gamma'(c^{\Gamma}) - G_{\pm}'(c_{\pm})),
\end{align}
\end{subequations}
where the density $\rho(\varphi)$ and viscosity $\eta(\varphi)$ are defined as
\begin{align}\label{rhovarphi}
\rho(\varphi) := \frac{\tilde{\rho}_{+} - \tilde{\rho}_{-}}{2} \varphi + \frac{\tilde{\rho}_{+} + \tilde{\rho}_{-}}{2}, \quad \eta(\varphi) :=  \frac{\eta_{+} - \eta_{-}}{2} \varphi + \frac{\eta_{+} + \eta_{-}}{2}.
\end{align}
Equations \eqref{PFMA:incompress} and \eqref{PFMA:momentum} are the incompressibility condition and the phase field momentum equations, respectively.  Equation \eqref{PFMA:phase} together with \eqref{PFMA:chem} forms a Cahn--Hilliard type equation which governs how the order parameter evolves and equations \eqref{PFMA:bulk} and \eqref{PFMA:interface} are the bulk and interfacial surfactant equations, respectively.  In \eqref{PFMA:chem}, the variable $\mu$ is often denoted as the chemical potential, and in \eqref{PFMA:phase}, $m \geq 0$ denotes a mobility for $\varphi$.  The above model \eqref{GLS:ModelA} is derived by modifying the approach of Teigen et al. \cite{Teigen:SolSurf} for the surfactant subsystem such that the following energy inequality is obtained (under suitable boundary conditions):
\begin{equation}
\label{PFM:Energy:Ineq}
\begin{aligned}
0 & = \frac{d}{dt} \int_{\Omega}  \left (  \sum \xi_{\pm}(\varphi) G_{\pm}(c_{\pm}) + \delta(\varphi, \nabla \varphi) \gamma(c^{\Gamma}) + \frac{\rho(\varphi)}{2} \abs{\vec{v}}^{2} \right ) \\
&  + \int_{\Omega} \left ( m(\varphi) \abs{\nabla \mu}^{2} + 2 \eta(\varphi) \abs{\vec{D} \vec{v}}^{2} + M_{\Gamma} \delta(\varphi, \nabla \varphi) \abs{\nabla \gamma'(c^{\Gamma})}^{2} \right ) \\
& + \int_{\Omega}  \sum \left ( M_{c}^{\pm}  \xi_{\pm}(\varphi) \abs{\nabla G_{\pm}'(c_{\pm})}^{2} + \frac{\delta(\varphi, \nabla \varphi)}{\alpha_{\pm}}  \abs{\gamma'(c^{\Gamma}) - G_{\pm}'(c_{\pm})}^{2} \right ).
\end{aligned}
\end{equation}  
Here, we observe the similarities between \eqref{SIM:Energy:Ineq} and \eqref{PFM:Energy:Ineq}.  In particular, $\delta(\varphi, \nabla \varphi) \gamma(c^{\Gamma})$ can be seen as an approximation of the interfacial surfactant energy density.

In the case of instantaneous adsorption for both fluid phases, that is, when the sub-layers in both bulk phases are in equilibrium with the interface, the ansatz is to assume that the chemical potentials $\gamma'(c^{\Gamma})$ and $G_{\pm}'(c_{\pm})$ are equal on the interface.  We can introduce the chemical potential as a new continuous variable $q$ and consider this as an unknown field and define the surfactant densities $c_{\pm}, c^{\Gamma}$ as functions of $q$:
\begin{align}\label{ModelC:ansatz}
c_{\pm}(q) := (G_{\pm}')^{-1}(q), \quad c^{\Gamma}(q) := (\gamma')^{-1}(q),
\end{align}
for strictly convex free energies $G_{\pm}$ and $\gamma$.  The surfactant densities are well-defined as the derivatives $G_{\pm}'$ and $\gamma'$ are monotone and one-to-one. Then, summing \eqref{PFMA:bulk} and \eqref{PFMA:interface} leads to one equation  for $q$:
\begin{align}\label{ModelC:surfactantequ}
\partial_{t}^{\bullet} \left ( \xi_{-} c_{-}(q) + \xi_{+} c_{+}(q) + \delta c^{\Gamma}(q) \right ) = \Div \, \left ( \left (M_{c}^{-} \xi_{-} + M_{c}^{+} \xi_{+} + M_{\Gamma} \delta \right ) \nabla q \right ) .
\end{align}
We define the surface tension $\tilde{\sigma}(q)$ by
\begin{align}\label{tilde:sigma}
\tilde{\sigma}(q) := \sigma(c^{\Gamma}(q)) = \gamma(c^{\Gamma}(q)) - q c^{\Gamma}(q).
\end{align}
Then the diffuse interface model for soluble surfactants of \cite{GarckeLamStinner} (denoted as Model C) in the case of instantaneous adsorption consists of \eqref{PFMA:incompress}-\eqref{PFMA:chem} and \eqref{ModelC:surfactantequ} (with $\tilde{\sigma}$ replacing $\sigma$ in \eqref{PFMA:momentum} and \eqref{PFMA:chem}, and $q$ replacing $G_{\pm}'(c_{\pm})$ in \eqref{PFMA:chem}).

It is also possible to consider a model which has instantaneous adsorption in $\Omega_{+}$ and dynamic adsorption in $\Omega_{-}$.  In this case we use \eqref{eq:instadsorption} to express $c^{\Gamma}$ as a function of $c_{+}$, and add \eqref{PFMA:interface} to the equation \eqref{PFMA:bulk} for $c_{+}$.  This yields a equation in $c_{+}$ that is coupled to the equation of $c_{-}$ via a source term $\frac{1}{\alpha} \delta (G_{+}'(c_{+}) - G_{-}'(c_{-}))$.  This is denoted as Model B in \cite{GarckeLamStinner}.

In \cite{GarckeLamStinner}, for the choice of a degenerate mobility $m(\varphi) = (1-\varphi^{2})_{+} = \max(0, 1-\varphi^{2})$, it has been shown via the method of formally matched asymptotic expansions that the sharp interface model \eqref{GLS:SIM} with $\alpha_{\pm} > 0$ is recovered from Model A in the limit $\eps \to 0$, and analogously \eqref{GLS:SIM} with \eqref{eq:instadsorption} instead of \eqref{SIM:dynamicAdsorp} is recovered from both Model C and Model A with the particular scaling $\alpha_{\pm} = \eps$.  We point out that the same sharp interface models can be recovered from \eqref{GLS:ModelA} if we consider the choice $m(\varphi) = \eps m_{0}$ for some positive constant $m_{0} > 0$.

In terms of the mathematical analysis of the aforementioned diffuse interface models, the main difficulty lies in getting a compactness result for the surfactant densities.  Take for example Model C with constant mobilities $M_{c}^{\pm} = M_{\Gamma} = 1$ and equal bulk energy densities $G_{-} = G_{+}$ (and hence $c_{-}(q) = c_{+}(q) =: c(q)$).  Then, Model C admits an energy identity of the form
\begin{equation}
\label{ModelC:Energy}
\begin{aligned}
0 & = \frac{d}{dt} \int_{\Omega}  \left ( G(c(q)) + \delta(\varphi, \nabla \varphi) \gamma(c^{\Gamma}(q)) + \frac{\rho(\varphi)}{2} \abs{\vec{v}}^{2} \right ) \\
&  + \int_{\Omega} \left ( m(\varphi) \abs{\nabla \mu}^{2} + 2 \eta(\varphi) \abs{\vec{D} \vec{v}}^{2} + \left ( 1 + \delta(\varphi, \nabla \varphi) \right ) \abs{\nabla q}^{2} \right ),
\end{aligned}
\end{equation}  
where we used $\xi_{-} + \xi_{+} = 1$.  If $\gamma$ is bounded from below by a positive constant, then one obtains spatial estimates for $\varphi$ in $H^{1}(\Omega)$, and compactness with respect to time follows from standard arguments.  However, any time compactness for $q$ has to come from equation \eqref{ModelC:surfactantequ}, which now reads as
\begin{align*}
\partial_{t}^{\bullet} \left ( c(q) + \delta(\varphi, \nabla \varphi) c^{\Gamma}(q) \right ) = \Div \, \left ( \left ( 1 + \delta(\varphi, \nabla \varphi) \right ) \nabla q \right ).
\end{align*}
This is not a trivial matter as $\abs{\nabla \varphi}^{2}$ appears under the time derivative, and thus compactness with respect to the strong topologies for $\nabla \varphi$ has to be derived beforehand.  The appearance of $\abs{\nabla \varphi}^{2}$ under the time derivative comes from the fact that we used
\begin{align}\label{original:Interfacial}
\int_{\Omega} \gamma(c^{\Gamma}(q)) \left ( \frac{\eps}{2} \abs{\nabla \varphi}^{2} + \frac{1}{\eps} \psi(\varphi) \right ) 
\end{align}
as an approximation to the interfacial surfactant energy.  An alternative is to model the interfacial surfactant energy with the help of the functional
\begin{align}
\label{ModelD:Interfacial}
\int_{\Omega} \left ( \frac{\eps}{2} \abs{\nabla \varphi}^{2} + \frac{d(q)}{\eps} \psi(\varphi) \right ), \text{ with }
d(q) := h(q) - h'(q) q, \enspace h(q) := (\tilde{\sigma}(q))^{2},
\end{align}
i.e., $h$ is the square of the surface tension $\tilde{\sigma}$ and $d$ is the Legendre transform of the square of the surface tension.  The difference between the original approximation \eqref{original:Interfacial} and the alternate approximation \eqref{ModelD:Interfacial} is that there are no functions involving $q$ that are multiplied with $\abs{\nabla \varphi}^{2}$.  Heuristically, we have transferred the interfacial surfactant energy from the gradient part all onto the potential part.  It turns out that the correct prefactor in front of the potential part is the Legendre transform of the square of the surface tension if we want to recover the appropriate sharp interface model.  Consequently, following the derivation in \cite{AbelsGarckeGrun,GarckeLamStinner}, we obtain the model (denoted as Model D hereafter)
\begin{subequations}\label{AGW}
\begin{align}
\Div \, \vec{v} & = 0, \label{ModelD:incompress} \\
\partial_{t}(\rho \vec{v})  + \Div \, ( \rho \vec{v} \otimes \vec{v} ) & = \Div \, \left ( - p Id + 2 \eta \vec{D} \vec{v} + \vec{v} \otimes \tfrac{\tilde{\rho}_{+} - \tilde{\rho}_{-}}{2} m(\varphi) \nabla \mu \right ) \label{ModelD:momentum} \\
\nonumber & + \Div \,  \left (  \left ( \frac{\eps}{2} \abs{\nabla \varphi}^{2} + \frac{(\tilde{\sigma}(q))^{2}}{\eps}  \psi(\varphi) \right ) Id - \eps \nabla \varphi \otimes \nabla \varphi \right ), \\
\partial_{t}^{\bullet} \varphi & =  \Div \, (m(\varphi) \nabla \mu),  \label{ModelD:phase} \\ 
\mu - \frac{(\tilde{\sigma}(q))^{2}}{\eps} \psi'(\varphi) & = - \eps \Delta \varphi + \sum \xi'_{\pm}(\varphi)(G_{\pm}(c_{\pm}(q)) - q c_{\pm}(q)), \label{ModelD:chem}  \\ 
\partial_{t}^{\bullet} \bigg{(} \frac{2}{\eps} \psi(\varphi) \tilde{\sigma}(q) c^{\Gamma}(q)  & +  \xi_{-}(\varphi) c_{-}(q) + \xi_{+}(\varphi) c_{+}(q)  \bigg{)} \label{ModelD:surfactant} \\
\nonumber & = \Div \, \left ( \left ( M_{c}^{-} \xi_{-}(\varphi) + M_{c}^{+} \xi_{+}(\varphi) +  \frac{2}{\eps} M_{\Gamma} \tilde{\sigma}(q) \psi(\varphi) \right ) \nabla q \right ). 
\end{align}
\end{subequations}
Let us point out the main differences between \eqref{AGW} and Model C.  For the equation involving the chemical potential $\mu$, the surface tension is only paired with $\psi'(\varphi)$.  This is also the case in the momentum equation.  Meanwhile, in the surfactant equation, the prefactor $\delta(\varphi, \nabla \varphi) = \frac{\eps}{2} \abs{\nabla \varphi}^{2} + \frac{1}{\eps} \psi(\varphi)$ is replaced by $\frac{2}{\eps} \psi(\varphi) \tilde{\sigma}(q)$.  Unlike in the previous models where the surface tension appears as a common factor in both the gradient term and the potential term, in this new model, we have transferred the prefactors all onto the potential term.  As we will discuss in Section \ref{sec:Asymptotics}, this causes the interfacial thickness to depend on the chemical potential $q$ (specifically see \eqref{Phi0}).  We point out that a similar situation also occurs when the interfacial energy depends on the orientation of the interface, see Garcke, Nestler and Stoth \cite{GNS} for example.

Under suitable boundary conditions, Model D \eqref{AGW} admits the following energy identity
\begin{align*}
 0 & = \frac{d}{dt} \int_{\Omega} \left ( \frac{\rho(\varphi)}{2} \abs{\vec{v}}^{2} + \frac{\eps}{2} \abs{\nabla \varphi}^{2} + \frac{d(q)}{\eps} \psi(\varphi) + \sum \xi_{\pm}(\varphi) G_{\pm}(c_{\pm}(q)) \right )  \\
  &+ \int_{\Omega} \left ( 2 \eta(\varphi) \abs{\vec{D} \vec{v}}^{2} + m(\varphi) \abs{\nabla \mu}^{2} + \left ( \sum M_{c}^{\pm} \xi_{\pm}(\varphi) + \frac{2}{\eps} M_{\Gamma} \tilde{\sigma}(q) \psi(\varphi) \right ) \abs{\nabla q}^{2} \right ) ,
\end{align*}
where $d(q)$ is defined in \eqref{ModelD:Interfacial}.  The above energy identity will be useful to show the existence of weak solutions to a simplified version of \eqref{AGW} in Section \ref{sec:Existence} below.   For numerical computations based on the models discussed in this section, we refer to \cite{AHKN,GarckeLamStinner}.

\section{Sharp interface limit}\label{sec:Asymptotics}

The sharp interface limit of diffuse interface models can be derived with the method of formally matched asymptotic expansions, which is described in detail in \cite{AbelsGarckeGrun,GarckeLamStinner,GStinner}.  In this section, we derive the sharp interface limit for Model D.  The procedure is similar to the one performed for Model C in \cite{GarckeLamStinner}, and so, in the following we only give a brief overview of the analysis.  We make the following assumptions:
\begin{align*}
& \tilde{\sigma}(s) > 0 \quad  \forall s \in \mathbb{R}, \quad \psi(\pm 1) = \psi'(\pm 1) = 0, \quad \psi(s) > 0 \quad \forall s \neq \pm 1, \\
& \xi_{-}(1) = 0, \; \xi_{-}(-1) = 1, \;  \xi_{+} = 1 - \xi_{-}, \quad m(\varphi) = \eps m_{0},
\end{align*}
where $m_{0} > 0$ is a fixed constant.  The idea of the method is as follows:  We assume that for small $\eps$, the domain $\Omega$ can be divided into two open subdomains $\Omega_{\pm}(t;\eps)$ at each time, separated by an interface $\Gamma(t;\eps)$.  There exists a family of solutions $(\varphi^{\eps}, \mu^{\eps}, \vec{v}^{\eps}, p^{\eps}, q^{\eps})$ to \eqref{AGW}, sufficiently smooth and indexed by $\eps$ such that the solutions have asymptotic expansions in $\eps$ in the bulk regions (away from $\Gamma(t;\eps)$ which are denoted as outer expansions) and another set of expansions in the interfacial regions (close to $\Gamma(t;\eps)$ which are denoted as inner expansions).  The idea is to analyse these expansions order by order in suitable transition regions where they should match up.  

For convenience, we define the flux
\begin{align}\label{asym:flux}
\bm{F} := \left ( \sum M_{c}^{\pm} \xi_{\pm}(\varphi) + \frac{2}{\eps} M_{\Gamma} \tilde{\sigma}(q) \psi(\varphi) \right ) \nabla q.
\end{align}

\paragraph{Outer expansions.} For $u^{\eps} = u(t,x;\eps) \in \{\varphi^{\eps}, \mu^{\eps}, \vec{v}^{\eps}, p^{\eps}, q^{\eps}\}$ we assume the following asymptotic expansion exists:
\begin{align*}
u^{\eps}(t,x) = u_{0}(t,x) + \eps u_{1}(t,x) + \text{ h.o.t.},
\end{align*}
where $\text{h.o.t.}$ denotes terms of higher order in $\eps$.  Due to the definition of the flux $\bm{F}$, we assume it has an outer expansion of the form
\begin{align*}
F^{\eps}(t,x) = \frac{1}{\eps^{2}} \bm{F}_{-2}^{b} + \frac{1}{\eps} \bm{F}_{-1}^{b} + \bm{F}_{0}^{b} + \text{ h.o.t.},
\end{align*}
where for instance
\begin{align*}
\bm{F}_{-2}^{b} = \bm{0}, \quad \bm{F}_{-1}^{b} = 2 \tilde{\sigma}(q_{0}) \psi(\varphi_{0}) M_{\Gamma}(c^{\Gamma}(q_{0})) \nabla q_{0}.
\end{align*}
To leading order \eqref{ModelD:chem} gives
\begin{align*}
h(q_{0}) \psi'(\varphi_{0}) = (\tilde{\sigma}(q_{0}))^{2} \psi'(\varphi_{0}) = 0.
\end{align*}
Since $h > 0$, this implies that $\psi'(\varphi_{0}) = 0$ and we choose $\varphi_{0}$ to be the stable minima of $\psi$, which are $\pm 1$.  This allows us to define (suppressing the the dependence on time)
$\Omega_{-} := \{ x \in \Omega: \varphi_{0}(x) = -1 \}$ and $\Omega_{+} := \{ x \in \Omega  : \varphi_{0}(x) = 1\}$ as the bulk fluid regions.  Then, as $\varphi_{0} = \pm 1$ and $\psi(\pm 1) = 0$, the term $\Div \,  ((\frac{\eps}{2} \abs{\nabla \varphi}^{2} + \frac{(\tilde{\sigma}(q))^{2}}{\eps} \psi(\varphi) ) Id )$ on the right-hand side of  \eqref{ModelD:momentum} does not contribute to leading order.  Furthermore, as the mobility is scaled with $\eps$, the terms involving $m$ also do not contribute to leading order.  Hence, we obtain from \eqref{ModelD:incompress} and \eqref{ModelD:momentum} the incompressible Navier--Stokes equations \eqref{SIM:incompress}-\eqref{SIM:momentum} in $\Omega_{\pm}$.  As $\psi(\pm 1) = 0$ and so $\bm{F}_{-1}^{b} = \bm{0}$, to leading order \eqref{ModelD:surfactant} yields a trivial identity.  Note that $\psi(\pm 1) = \psi'(\pm 1) = 0$ implies that
\begin{align}\label{F0b}
\bm{F}_{0}^{b} = \sum \xi_{\pm}(\varphi_{0}) M_{c}^{\pm}(c_{\pm}(q_{0})) \nabla q_{0},
\end{align}  
and so to first order we obtain from \eqref{ModelD:surfactant}
\begin{align*}
& \partial_{t}^{\bullet} \left ( \xi_{-}(\varphi_{0}) c^{-}(q_{0}) + \xi_{+}(\varphi_{0}) c^{+}(q_{0}) \right ) = \Div  \, \bm{F}_{0}^{b} \\
& \quad = \Div \, \left ( \xi_{-}(\varphi_{0}) M_{c}^{-} \nabla q_{0} + \xi_{+}(\varphi_{0}) M_{c}^{+} \nabla q_{0} \right ),
\end{align*}
where we have used $\psi(\pm 1) = \psi'(\pm 1) = 0$.  Then, using the properties of $\xi_{\pm}$ we obtain
\begin{align}
\partial_{t}^{\bullet} c_{\pm}(q_{0}) = \Div \, (M_{c}^{\pm}(c_{\pm}(q_{0})) \nabla q_{0}) \text{ in } \Omega_{\pm}.
\end{align}
\paragraph{Inner expansions and matching conditions.}   We assume that the zero level sets of $\varphi_{\eps}$ converge to some limiting hypersurface $\Gamma$ moving with a normal velocity $u_{\Gamma}$ as $\eps \to 0$.  Let $d(t,x)$ denote the signed distance function to $\Gamma$ with the convention $d(t,x) > 0$ for $x \in \Omega_{+}$, and setting $z(t,x) = d(t,x)/\eps$ as the rescaled signed distance function, we can express functions $u(t,x)$ close to $\Gamma$ in a new coordinate system as $U(t,s,z)$, where $s$ denotes the tangential spatial coordinates on $\Gamma$.  Introducing the normal time derivative $\partial_{t}^{\circ}(\cdot) = \partial_{t}(\cdot) + u_{\Gamma} \cdot \nabla (\cdot)$, one obtains the following transformations
\begin{align*}
\partial_{t}u = -\frac{1}{\eps} u_{\Gamma} \partial_{z}U + \partial_{t}^{\circ} U + \text{ h.o.t.}, \quad \nabla_{x} u = \frac{1}{\eps} \bm{\nu} \partial_{z}U + \nabla_{\Gamma} U + \text{ h.o.t.},
\end{align*}
where $\bm{\nu} = \nabla_{x} d$ is the unit normal pointing into $\Omega_{+}$, and $\nabla_{\Gamma}$ is the tangential gradient on $\Gamma$.  The inner expansions of $u^{\eps} \in \{\varphi^{\eps}, \mu^{\eps}, \vec{v}^{\eps}, p^{\eps}, q^{\eps} \}$ takes the form
\begin{align*}
u^{\eps}(t,x) = U(t,s,z;\eps) = U_{0}(t,s,z) + \eps U_{1}(t,s,z) + \text{ h.o.t.},
\end{align*}
with corresponding inner variables $U \in \{ \Phi, \Xi, \vec{V}, P, Q\}$.  For the flux $\bm{F}$ \eqref{asym:flux} we assume the following inner expansion
\begin{align*}
\bm{F}_{\eps}(t,x) = \frac{1}{\eps^{2}} \bm{F}_{-2}^{i}(t,s,z) + \frac{1}{\eps} \bm{F}_{-1}^{i}(t,s,z) + \bm{F}_{0}^{i}(t,s,z) + \text{ h.o.t.},
\end{align*}
where for example
\begin{align}\label{Flux:inner:-2:term}
\bm{F}_{-2}^{i}(t,s,z) = 2 \tilde{\sigma}(Q_{0}) \psi(\Phi_{0}) M_{\Gamma}(c^{\Gamma}(Q_{0})) \partial_{z}Q_{0} \bm{\nu}.
\end{align}
We further assume that $\Phi_{0}(t,s,0) = 0$, which arises from the assumption that the zero level set of $\varphi_{\eps}$ converge to $\Gamma$.  In order to match the inner expansions valid in the interfacial region to the outer expansions we employ the following matching conditions \cite{GStinner}:
\begin{equation*}
\begin{alignedat}{5}
&\lim_{z \to \pm \infty} U_{0}(t,s,z) = u_{0}^{\pm}(t,x), \quad && \lim_{z \to \pm \infty} \partial_{z}U_{0}(t,s,z) = 0, \quad && \lim_{z \to \pm \infty} \partial_{z}U_{1}(t,s,z) = \nabla u_{0}^{\pm} \cdot \bm{\nu}, \\
&\lim_{z to \pm \infty} \bm{F}_{-2}^{i}(t,s,z)  = \bm{0}, \quad && \lim_{z \to \pm \infty} \partial_{z} \bm{F}_{-2}^{i}(t,s,z) = 0, \quad && \lim_{z \to \pm \infty} \bm{F}_{-1}^{i}(t,s,z) = \bm{0}, \\
&\lim_{z \to \pm \infty} \bm{F}_{0}^{i}(t,s,z)  = (\bm{F}_{0}^{b})^{\pm}(t,x),&& &&
\end{alignedat}
\end{equation*}
where $u_{0}^{\pm}(t,x) := \lim_{\delta \to 0} u_{0}(t, x \pm \delta \bm{\nu}(x))$ for $x \in \Gamma$ such that $x + \delta \bm{\nu}(x) \in \Omega_{+}$ and $x + \delta \bm{\nu}(x) \in \Omega_{-}$, and we have used that $\bm{F}_{-2}^{b} = \bm{F}_{-1}^{b} = \bm{0}$.  Then, to leading order \eqref{ModelD:surfactant} yields $\partial_{z} \bm{F}_{-2}^{i} \cdot \bm{\nu} = \partial_{z}(\bm{F}_{-2}^{i} \cdot \bm{\nu}) = 0$, which implies that $\bm{F}_{-2}^{i} \cdot \bm{\nu}$ is constant in $z$.  For any tangential vector $\bm{\tau}$, we have by definition \eqref{Flux:inner:-2:term} that $\bm{F}_{-2}^{i} \cdot \bm{\tau} = 0$.  Thus, by the matching conditions we obtain that $\bm{F}_{-2}^{i} \equiv \bm{0}$, which in turn implies that
\begin{align}\label{pdzQ0zero}
\partial_{z} Q_{0} = 0 \text{ whenever } \abs{\Phi_{0}} < 1.
\end{align}
To leading order \eqref{ModelD:chem} yields
\begin{align}\label{ODE}
\partial_{zz} \Phi_{0} - (\tilde{\sigma}(Q_{0}))^{2} \psi'(\Phi_{0}) = 0.
\end{align}
We consider the function $\phi(z)$ satisfying
\begin{align}\label{phi:property}
\phi''(z) = \psi'(\phi(z)), \quad \lim_{z \to \pm \infty} \phi(z) = \pm 1, \quad \phi(0) = 0.
\end{align}
For the double-well potential $\psi(s) = \frac{1}{4}(1-s^{2})^{2}$, the solution is $\phi(z) = \tanh(z/\sqrt{2})$ whose derivative satisfies
\begin{align}\label{expo:decay}
\lim_{t \to \pm \infty} t \abs{\phi'(t)}^{2} = 0.
\end{align}
We now set 
\begin{align}\label{Phi0}
\Phi_{0}(t,s,z) = \phi(\tilde{\sigma}(Q_{0}(t,s))z).
\end{align}
A short calculation shows that $\Phi_{0}$ indeed solves \eqref{ODE} with $\Phi(t,s,0) = 0$.  Here we point out that, in the asymptotic analysis of Model A and Model C, $\Phi_{0}$ is a function depending only on $z$, and so \eqref{ODE} is a new feature of Model D, which states that the interfacial thickness depends on $q$.  Multiplying \eqref{ODE} with $\partial_{z}\Phi_{0}$, integrating over $z$ and applying the matching conditions leads to the equipartition of energy:
\begin{equation}\label{equipartition}
\begin{aligned}
& \frac{1}{2} \abs{\partial_{z}\Phi_{0}}^{2}(t,s,z) = (\tilde{\sigma}(Q_{0}))^{2}(t,s) \psi(\Phi_{0})(t,s,z),\\
& \quad \text{ with } \int_{\mathbb{R}} \abs{\partial_{z}\Phi_{0}}^{2}(t,s,z) \, dz = (\tilde{\sigma}(q_{0}))^{2} \int_{\mathbb{R}} 2 \psi(\phi(\tilde{\sigma}(q_{0}) z) \, dz = \tilde{\sigma}(q_{0}),
\end{aligned}
\end{equation}
where for the last equality a change of variables $t \mapsto \tilde{\sigma}(q_{0})z$ and the fact that $\psi$ is rescaled so that $\int_{\mathbb{R}} 2 \psi(\phi(t)) \, dt = \int_{-1}^{1} \sqrt{2 \psi(s)} \, ds = 1$ are used.   Using \eqref{ModelC:ansatz}, \eqref{tilde:sigma} and \eqref{ModelD:Interfacial}, we obtain the relations:
\begin{align*}
h'(q) = 2 \tilde{\sigma}(q) \tilde{\sigma}'(q), \enspace \tilde{\sigma}'(q) = \gamma'(c^{\Gamma}(q)) (c^{\Gamma})'(q) - c^{\Gamma}(q) - q (c^{\Gamma})'(q) = -c^{\Gamma}(q).
\end{align*} 
Using \eqref{ModelD:Interfacial}, this leads to the correct formula for the total energy across the interface:
\begin{align*}
\int_{\mathbb{R}} \frac{1}{2} |\partial_{z}\Phi_{0}|^{2} + d(q_{0}) \psi(\phi(\tilde{\sigma}(q_{0})z)) \, dz = \tilde{\sigma}(q_{0}) - \tilde{\sigma}'(q_{0}) q_{0} = \gamma(c^{\Gamma}(q_{0})).
\end{align*}
Meanwhile, to leading order, we obtain from \eqref{ModelD:incompress}, \eqref{ModelD:phase} that
\begin{align*}
\partial_{z} \bm{V}_{0} \cdot \bm{\nu} = 0, \quad (-u_{\Gamma} + \bm{V}_{0} \cdot \bm{\nu}) \partial_{z}\Phi_{0} = m_{0} \partial_{zz} \Xi_{0}.
\end{align*}
Integrating and applying matching conditions leads to $[\bm{v}_{0}]_{-}^{+} \cdot \bm{\nu} = 0$, and simultaneously $\bm{v}_{0} \cdot \bm{\nu} = u_{\Gamma}$ and $\partial_{z}\Xi_{0} = 0$ (see \cite{AbelsGarckeGrun} for more details).  Using $\partial_{z} Q_{0} = \partial_{z} \bm{V}_{0} \cdot \bm{\nu} = \partial_{z}\Xi_{0} = 0$ and \eqref{ODE}, to leading order \eqref{ModelD:momentum} yields $\bm{0} = \partial_{z} (\eta(\Phi_{0}) \partial_{z} \bm{V}_{0})$.  Integrating with respect to $z$ and applying the matching conditions yields $\partial_{z} \bm{V}_{0} = \bm{0}$ and hence $[\bm{v}_{0}]_{-}^{+} = \bm{0}$.  Next, using that $\bm{F}_{-2}^{i} = \bm{0}$, $\partial_{z} Q_{0} = 0$, $\nabla_{\Gamma} Q_{0} \cdot \bm{\nu} = 0$, $u_{\Gamma} = \bm{v}_{0} \cdot \bm{\nu}$ and that $\bm{\nu}$ is independent of $z$, to first order \eqref{ModelD:surfactant} gives
\begin{align*}
0 = \partial_{z} \bm{F}_{-1}^{i} \cdot \bm{\nu} = 2\partial_{z} ( \tilde{\sigma}(Q_{0}) \psi(\Phi_{0}) M_{\Gamma}(c^{\Gamma}(Q_{0})) \partial_{z} Q_{1}).
\end{align*}
Integrating with respect to $z$ and the properties of $\tilde{\sigma}$ and $\psi$ yield that 
\begin{align*}
\partial_{z} Q_{1} = 0 \text{ whenever } \abs{\Phi_{0}} < 1.
\end{align*}
To first order \eqref{ModelD:chem} gives
\begin{equation}\label{ModelD:chem:inner:first}
\begin{aligned}
& \Xi_{0} - \sum \xi_{\pm}'(\Phi_{0})(G_{\pm}(c_{\pm}(Q_{0})) - Q_{0} c_{\pm}(Q_{0})) \\
& \quad = h(Q_{0}) \psi''(\Phi_{0}) \Phi_{1} + h'(Q_{0})Q_{1} \psi'(\Phi_{0}) - \partial_{zz}\Phi_{1} + \partial_{z}\Phi_{0} \kappa,
\end{aligned}
\end{equation}
where we used that $\Div_{\Gamma} \, (\partial_{z}\Phi_{0} \bm{\nu}) = \partial_{z}\Phi_{0} \, \Div_{\Gamma}\, \bm{\nu} = - \partial_{z}\Phi_{0} \kappa$ with the mean curvature $\kappa$.  Multiplying \eqref{ModelD:chem:inner:first} with $\partial_{z}\Phi_{0}$, writing $f'(\Phi_{0}) \partial_{z}\Phi_{0} = \partial_{z} f(\Phi_{0})$ for $f \in \{ \xi_{\pm}, \psi'\}$, integrating with respect to $z$, integrating by parts for the right-hand side, using $\partial_{z}\Xi_{0} = \partial_{z}Q_{0} = \partial_{z} Q_{1} = 0$, \eqref{ODE}, \eqref{equipartition}, and applying the matching conditions and the properties of $\xi_{\pm}$, $\psi$ and $\psi'$ at $\pm 1$ gives
\begin{align*}
& 2 \mu_{0} - [G(c(q_{0})) - q_{0} c(q_{0})]_{-}^{+} - \tilde{\sigma}(q_{0}) \kappa = \int_{\mathbb{R}}  (\partial_{zz}\Phi_{0} - h(Q_{0}) \psi'(\Phi_{0})) \partial_{z}\Phi_{1} \, dz\\
& \quad + [h'(Q_{0}) Q_{1} \psi(\Phi_{0}) - \partial_{z}\Phi_{0} \partial_{z}\Phi_{1} + h(Q_{0}) \psi'(\Phi_{0}) \Phi_{1}]_{z=-\infty}^{z=+\infty} = 0.
\end{align*}
That is, we obtain $2 \mu_{0} = \tilde{\sigma}(q_{0}) \kappa + [G(c(q_{0})) - q_{0} c(q_{0})]_{-}^{+}$ as a solvability condition for $\Phi_{1}$.  To first order \eqref{ModelD:incompress} gives $\partial_{z} \bm{V}_{1} \cdot \bm{\nu} = - \Div_{\Gamma} \, \bm{V}_{0}$.  Furthermore, since $\partial_{z} \Xi_{0} = 0$ and $\partial_{z} \bm{V}_{0} = \bm{0}$, the term $\Div \, (\bm{v} \otimes m(\varphi) \nabla \mu)$ in the momentum equation \eqref{ModelD:momentum} does not contribute to the first order inner expansion.  Similarly, $u_{\Gamma} = \bm{v}_{0} \cdot \bm{\nu}$ implies that the left-hand side of \eqref{ModelD:momentum} also does not contribute.  Using $\partial_{z} \bm{V}_{0} = \bm{0}$ the first line of the right-hand side of \eqref{ModelD:momentum} gives to first order
\begin{align*}
-[p_{0} Id - 2 \eta \vec{D} \vec{v}_{0}]_{-}^{+} \bm{\nu}
\end{align*}
after integrating with respect to $z$ and matching, see \cite{AbelsGarckeGrun,GarckeLamStinner} for more details.  Meanwhile, for the second line on the right-hand side of \eqref{ModelD:momentum} we obtain to first order
\begin{align*}
& \partial_{z} \left ( -\partial_{z}\Phi_{0} \partial_{z} \Phi_{1} + h(Q_{0}) \psi'(\Phi_{0}) \Phi_{1} + h'(Q_{0})Q_{1} \psi(\Phi_{0}) \right ) \bm{\nu} - \partial_{z}(\partial_{z}\Phi_{0} \nabla_{\Gamma} \Phi_{0}) \\
& \quad + \nabla_{\Gamma} \left ( \frac{1}{2} \abs{\partial_{z}\Phi_{0}}^{2} +  h(Q_{0}) \psi(\Phi_{0}) \right ) - \Div_{\Gamma} \, (\abs{\partial_{z}\Phi_{0}}^{2} \bm{\nu} \otimes \bm{\nu}).
\end{align*}
Note that after integrating with respect to $z$ and applying the matching conditions, the first term vanishes.  Furthermore, it can be shown using a change of variables and \eqref{expo:decay} that
\begin{align*}
[\partial_{z}\Phi_{0} \nabla_{\Gamma} \Phi_{0}]_{z=-\infty}^{z=+\infty} = \nabla_{\Gamma} \tilde{\sigma}(q_{0}) [(\phi'(\tilde{\sigma}(q_{0})z))^{2} \tilde{\sigma}(q_{0})z]_{z=-\infty}^{z=+\infty} = \nabla_{\Gamma} \tilde{\sigma}(q_{0}) [t (\phi'(t))^{2}]_{t=-\infty}^{t=+\infty} = 0.
\end{align*}
Meanwhile, by the equipartition of energy \eqref{equipartition} we obtain
\begin{align*}
& \int_{\mathbb{R}} \nabla_{\Gamma} \left ( \frac{1}{2} \abs{\partial_{z}\Phi_{0}}^{2} +  h(Q_{0}) \psi(\Phi_{0}) \right ) - \Div_{\Gamma} \, (\abs{\partial_{z}\Phi_{0}}^{2} \bm{\nu} \otimes \bm{\nu}) \, dz \\
& \quad = \nabla_{\Gamma} \tilde{\sigma}(q_{0}) - \Div_{\Gamma} \, (\tilde{\sigma}(q_{0}) \bm{\nu} \otimes \bm{\nu}) = \nabla_{\Gamma} \tilde{\sigma}(q_{0}) + \tilde{\sigma}(q_{0}) \kappa \bm{\nu},
\end{align*}
and thus to first order we obtain from \eqref{ModelD:momentum} the condition
\begin{align*}
[p_{0} Id + 2 \eta^{(i)} \vec{D} \vec{v}_{0}]_{-}^{+} \bm{\nu} = \tilde{\sigma}(q_{0}) \kappa \bm{\nu} + \nabla_{\Gamma} \tilde{\sigma}(q_{0}).
\end{align*}
Using $u_{\Gamma} = \bm{v}_{0} \cdot \bm{\nu}$ and $\bm{F}_{-2}^{i} = \bm{0}$, to second order \eqref{ModelD:surfactant} gives
\begin{align*}
\partial_{t}^{\circ} g(\Phi_{0},Q_{0}) + \bm{V}_{0} \cdot \nabla_{\Gamma} g(\Phi_{0},Q_{0}) + \bm{V}_{1} \cdot \bm{\nu} \partial_{z}g(\Phi_{0},Q_{0}) = \partial_{z} \bm{F}_{0}^{i} \cdot \bm{\nu} + \Div_{\Gamma} \, \bm{F}_{-1}^{i},
\end{align*}
where we set $g(\Phi_{0},Q_{0}) = 2 \psi(\Phi_{0}) \tilde{\sigma}(Q_{0}) c^{\Gamma}(Q_{0})$. Furthermore, using that $\partial_{z}Q_{0} = \partial_{z}Q_{1} = 0$, we compute that
\begin{align*}
\bm{F}_{-1}^{i} = 2 \psi(\Phi_{0}) \tilde{\sigma}(Q_{0}) M_{\Gamma}(c^{\Gamma}(Q_{0})) \nabla_{\Gamma} Q_{0}. 
\end{align*} 
Then, integrating with respect to $z$ and applying the matching conditions, we have for the right-hand side
\begin{align*}
\bm{F}_{0}^{i} \vert_{z=-\infty}^{z=+\infty} \cdot \bm{\nu} + \Div_{\Gamma} \, \left ( \int_{\mathbb{R}} \bm{F}_{-1}^{i} \, dz \right ) = [M_{c}(c(q_{0})) \nabla q_{0}]_{-}^{+} \bm{\nu} + \Div_{\Gamma} \left ( M_{\Gamma}(c^{\Gamma}(q_{0})) \nabla_{\Gamma} q_{0} \right ) ,
\end{align*}
where we used the property that $\tilde{\sigma}(q_{0}) \int_{\mathbb{R}} 2 \psi(\Phi_{0}) \, dz = \int_{\mathbb{R}} 2 \psi(\phi(t)) \, dt = 1$ and the properties of $\xi_{\pm}(\pm 1)$.  Meanwhile, for the left-hand side, using integration by parts, the fact that $\partial_{z} \bm{V}_{1} \cdot \bm{\nu} = - \Div_{\Gamma} \, \bm{V}_{0}$ and the matching conditions leads to
\begin{align*}
& \partial_{t}^{\circ} c^{\Gamma}(q_{0}) + \bm{v}_{0} \cdot \nabla_{\Gamma} c^{\Gamma}(q_{0}) + \int_{\mathbb{R}} \bm{V}_{1} \cdot \bm{\nu} \partial_{z}g(\Phi_{0}, Q_{0}) \, dz \\
& \quad = \partial_{t}^{\circ} c^{\Gamma}(q_{0}) + \bm{v}_{0} \cdot \nabla_{\Gamma} c^{\Gamma}(q_{0}) + \int_{\mathbb{R}} (\Div_{\Gamma} \, \bm{V}_{0}) g(\Phi_{0}, Q_{0}) \, dz + [\bm{V}_{1} \cdot \bm{\nu} g(\Phi_{0}, Q_{0})]_{z=-\infty}^{z=+\infty} \\
& \quad = \partial_{t}^{\circ} c^{\Gamma}(q_{0}) + \bm{v}_{0} \cdot \nabla_{\Gamma} c^{\Gamma}(q_{0}) + c^{\Gamma}(q_{0}) \, \Div_{\Gamma} \, \bm{v}_{0} = \partial_{t}^{\bullet} c^{\Gamma}(q_{0}) + c^{\Gamma}(q_{0}) \, \Div_{\Gamma} \, \bm{v}_{0},
\end{align*}
where we have used that $\partial_{t}^{\bullet} (\cdot) = \partial_{t}^{\circ} (\cdot) + \bm{v} \cdot \nabla_{\Gamma} (\cdot)$, and the fact that $\int_{\mathbb{R}} g(\Phi_{0}, Q_{0}) \, dz = c^{\Gamma}(q_{0})$.  The jump term vanishes due to $\psi(\pm 1) = 0$.  Altogether we obtain the equation
\begin{align*}
\partial_{t}^{\bullet} c^{\Gamma}(q_{0}) + c^{\Gamma}(q_{0}) \, \Div_{\Gamma} \, \bm{v}_{0} = \Div_{\Gamma} \, (M_{\Gamma}(c^{\Gamma}(q_{0})) \nabla_{\Gamma} q_{0}) + [M_{c}(c(q_{0})) \nabla q_{0}]_{-}^{+} \bm{\nu}.
\end{align*}
Hence, the sharp interface model of Model D \eqref{AGW} is \eqref{SIM:incompress}, \eqref{SIM:momentum}, \eqref{SIM:velocityjump}, together with
\begin{equation*}
\begin{aligned}
\partial_{t}^{\bullet} c_{\pm}(q) & = \Div \, (M_{c}^{\pm}(c_{\pm}(q)) \nabla q) && \text{ in } \Omega_{\pm} (t), \\
[p Id - 2 \eta \bm{D} \bm{v}]_{-}^{+} \bm{\nu} & = \tilde{\sigma}(q) \kappa \bm{\nu} + \nabla_{\Gamma} \tilde{\sigma}(q) && \text{ on } \Gamma(t), \\
\partial_{t}^{\bullet} c^{\Gamma}(q) + c^{\Gamma}(q) \, \Div_{\Gamma}  \, \bm{v} - \Div_{\Gamma} \, (M_{\Gamma}(c^{\Gamma}(q)) \nabla_{\Gamma} q) & = [M_{c}(c(q)) \nabla q]_{-}^{+} \bm{\nu} && \text{ on } \Gamma(t).
\end{aligned}
\end{equation*}

\section{Existence result}\label{sec:Existence}
Setting $G_{-} = G_{+} =: G$, $M_{c}^{-} = M_{c}^{+} =: M_{c}$, and $c_{-} = c_{+} =: c$, the surfactant equation \eqref{ModelD:surfactant} can be expressed as
\begin{align}
\label{Josef:surfactant}
\partial_{t}^{\bullet} \left ( \frac{f(q)}{\eps} \psi(\varphi) + c(q) \right ) = \Div \, \left ( M(\varphi, q) \nabla q \right ),
\end{align}
where
\begin{align*}
f(q) := - h'(q) = 2 \tilde{\sigma}(q) c^{\Gamma}(q), \quad M(\varphi,q) := M_{c} +  \frac{2}{\eps} M_{\Gamma} \tilde{\sigma}(q) \psi(\varphi).
\end{align*}
In this section, let $T > 0$ be fixed and $\Omega \subset \mathbb{R}^{d}$, $d = 2,3$, be a bounded domain with sufficiently smooth boundary $\partial \Omega$.  Setting $Q_{T} := \Omega \times (0,T)$, we provide an existence result to the following model:
\begin{subequations}\label{Model:Josef}
\begin{alignat}{3}
\Div \, \vec{v} & = 0 && \text{ in } Q_{T}, \label{ModelJ:incompress} \\
\nonumber \partial_{t}(\rho \vec{v})  + \Div \, ( \vec{v} \otimes (\rho \vec{v} + \tilde{\vec{J}} )) & = - \nabla p + \Div \, \left ( 2 \eta \vec{D} \vec{v} \right ) && \\
& + \left ( \mu - \frac{h(q)}{\eps} \psi'(\varphi) \right ) \nabla \varphi  + \frac{1}{2} R \vec{v} && \text{ in } Q_{T}, \label{ModelJ:momentum} \\
\partial_{t}^{\bullet} \varphi & =  \Div \, (m(\varphi) \nabla \mu) && \text{ in } Q_{T},  \label{ModelJ:phase} \\
\mu  & = - \eps \Delta \varphi + \frac{h(q)}{\eps} \psi'(\varphi) && \text{ in } Q_{T}, \label{ModelJ:chem}  \\ 
\partial_{t}^{\bullet} \left ( \frac{f(q)}{\eps} \psi(\varphi) + c(q) \right ) & = \Div \, \left ( M(\varphi, q) \nabla q  \right ) && \text{ in } Q_{T}, \label{ModelJ:surfactant} 
\end{alignat}
\end{subequations}
where 
\begin{align*}
\tilde{\vec{J}} := - \rho'(\varphi) m(\varphi) \nabla \mu, \quad R := -m(\varphi) \nabla \rho'(\varphi) \cdot \nabla \mu,
\end{align*}
together with the initial conditions
\begin{equation}\label{ModelJ:Initial}
\begin{aligned}
\varphi(0) & = \varphi_{0}, \quad \vec{v}(0) = \vec{v}_{0}, \\
\frac{f(q(0))}{\eps} \psi(\varphi(0)) + c(q(0)) & = \frac{f(q_{0})}{\eps} \psi(\varphi_{0}) + c(q_{0}) \quad \text{ in } \Omega,
\end{aligned}
\end{equation}
and the boundary conditions
\begin{align}\label{ModelJ:Bdy}
\vec{v} = \vec{0}, \quad \partial_{n} \varphi = \partial_{n} \mu = \partial_{n} q = 0 \text{ on } \partial \Omega \times (0,T),
\end{align}
where $\partial_{n} f = \nabla f \cdot \bm{\nu}$ denotes the normal derivative of $f$ on $\partial \Omega$.

Note that \eqref{ModelJ:surfactant} is exactly \eqref{Josef:surfactant}, and by choosing the free energy $G$ such that $G(c(q)) = q c(q)$, the second term on the right-hand side of \eqref{ModelD:chem} vanishes, leading to \eqref{ModelJ:chem}.  Furthermore, in \eqref{ModelD:momentum} we have replaced $p - \frac{h(q)}{\eps} \psi(\varphi)$ by a rescaled pressure, which we call $p$ again, and used the relation
\begin{align*}
\Div \, \left ( \frac{\eps}{2} \abs{\nabla \varphi}^{2} Id - \eps \nabla \varphi \otimes \nabla \varphi \right ) = -\eps \Delta \varphi \nabla \varphi = \left ( \mu - \frac{h(q)}{\eps} \psi'(\varphi) \right ) \nabla \varphi.
\end{align*}    

As the density is a physical quantity that is positively valued, the explicit form \eqref{rhovarphi} for the density $\rho(\varphi)$ may become negative for certain values of $\varphi$, and in general, it is not guaranteed that the values of the order parameter will not deviate from the physical interval $[-1,1]$.  Hence, for the mathematical analysis of the models, the expression \eqref{rhovarphi} has to be modified in such a way that $\rho(s) > 0$ for all $s \in \mathbb{R}$, and this modification leads to the appearance of the terms $\tilde{\vec{J}}$ and $R\vec{v}$ in the momentum equation \eqref{ModelJ:momentum}.  Furthermore, in the physical interval $\varphi \in [-1,1]$, it holds that $\rho'(\varphi) = \frac{1}{2}(\tilde{\rho}_{+} - \tilde{\rho}_{-})$, and thus $\tilde{\vec{J}} = -\frac{1}{2}(\tilde{\rho}_{+} - \tilde{\rho}_{-}) m(\varphi) \nabla \mu$, while $R = 0$.  Then, the corresponding momentum equation \eqref{ModelJ:momentum} is identical to \eqref{ModelD:momentum} (with a rescaled pressure).

To state the existence results, we introduce some notation and function spaces.  For $\vec{a}, \vec{b} \in \mathbb{R}^{d}$, the tensor product $\vec{a} \otimes \vec{b}$ is defined as $\vec{a} \otimes \vec{b} := (a_{i} b_{j})_{i,j=1}^{d}$.  If $\vec{A}, \vec{B} \in \mathbb{R}^{d \times d}$, then we set $\vec{A} : \vec{B} := \sum_{i,j=1}^{d} A_{ij} B_{ij}$.  Let $\Omega \subset \mathbb{R}^{d}$ be a bounded domain with $C^{2}$-boundary $\partial \Omega$.  For $1 \leq p \leq \infty$ and $k \in \mathbb{N} \cup \{0\}$, we denote by $L^{p}(\Omega)$ and $W^{k,p}(\Omega)$ the usual Lebesgue and Sobolev spaces equipped with the norms $|| \cdot ||_{L^{p}}$ and $|| \cdot ||_{W^{k,p}}$, respectively.  In the case $p = 2$, we use the notation $H^{k}(\Omega) := W^{k,2}(\Omega)$ for $k \geq 1$, along with the norm $|| \cdot ||_{H^{k}}  := || \cdot ||_{W^{k,2}}$.  We denoted $C^{\infty}_{0,\sigma}(\Omega) := \{ \vec{u} \in C^{\infty}_{0}(\Omega)^{d} : \Div \, \vec{u} = 0 \}$, and define $L^{2}_{\sigma}(\Omega)$ as the completion of $C^{\infty}_{0,\sigma}(\Omega)$ with respect to the $|| \cdot ||_{L^{2}}$ norm.  Furthermore, we define the space $H^{1}_{0}(\Omega)$ as the completion of $C^{\infty}_{0}(\Omega)$ with respect to the $|| \cdot ||_{H^{1}}$ norm, and use the notation $H^{2}_{n}(\Omega) := \{ f \in H^{2}(\Omega) : \partial_{n} f = 0 \text{ on } \partial \Omega \}$.

\begin{defn}[Weak solution]\label{defn:weaksoln}
Let $T \in (0,\infty)$, $\vec{v}_{0} \in L^{2}_{\sigma}(\Omega)$, $\varphi_{0} \in H^{2}_{n}(\Omega)$, and $q_{0} \in L^{2}(\Omega)$ be given.  We call $(\vec{v}, \varphi, \mu, q)$ a weak solution of \eqref{Model:Josef}-\eqref{ModelJ:Bdy} if
\begin{equation*}
\begin{alignedat}{5}
\vec{v} & \in L^{2}(0,T;H^{1}_{0}(\Omega)^{d}) \cap L^{\infty}(0,T;L^{2}_{\sigma}(\Omega)), \; && q && \in L^{2}(0,T;H^{1}(\Omega)) \cap L^{\infty}(0,T;L^{2}(\Omega)), \\
\varphi & \in L^{\infty}(0,T;H^{1}(\Omega)) \cap L^{2}(0,T;H^{2}(\Omega)), \quad && \mu && \in L^{2}(0,T;H^{1}(\Omega)),
\end{alignedat}
\end{equation*}
and the following equations are satisfied:
\begin{equation}
\begin{aligned}
& \int_{Q_{T}} - \rho(\varphi) \vec{v} \cdot \partial_{t} \vec{w} - (\rho(\varphi) \vec{v} \otimes \vec{v} + \vec{v} \otimes \tilde{\vec{J}}) : \nabla \vec{w} + 2 \eta(\varphi) \vec{D} \vec{v} : \vec{D} \vec{w} \, dx \, dt \\
& + \int_{Q_{T}} \frac{1}{2} m(\varphi) \left ( \nabla \rho'(\varphi) \cdot \nabla \mu \right ) \vec{v} \cdot \vec{w} - \left ( \mu - \frac{h(q)}{\eps} \psi'(\varphi) \right ) \nabla \varphi \cdot \vec{w} \, dx \, dt = 0
\end{aligned}
\end{equation}
for all $\vec{w} \in C^{\infty}_{0}(0,T;C^{\infty}_{0,\sigma}(\Omega))$ and
\begin{align}
\int_{Q_{T}} M(\varphi, q) \nabla q \cdot \nabla \xi - \left ( \frac{1}{\eps} f(q) \psi(\varphi) + c(q) \right ) \partial_{t}^{\bullet} \xi \, dx \, dt & = 0, \\
\int_{Q_{T}} m(\varphi) \nabla \mu \cdot \nabla \xi - \varphi \partial_{t} \xi + \nabla \varphi \cdot \vec{v} \xi \, dx \, dt & = 0, \\
\int_{Q_{T}} \mu \xi - \eps \nabla \varphi \cdot \nabla \xi - \frac{h(q)}{\eps} \psi'(\varphi) \xi \, dx \, dt & = 0
\end{align}
for all $\xi \in C^{\infty}_{0}(0,T;C^{1}(\overline{\Omega}))$.  Moreover, the energy inequality
\begin{align}
E(t) + \int_{s}^{t} \int_{\Omega} \left ( M(\varphi, q) |\nabla q|^{2} + m(\varphi) |\nabla \mu|^{2} + 2 \eta(\varphi) | \vec{D} \vec{v} |^{2} \right ) \, dx \, d \tau \leq E(s)
\end{align}
has to hold for all $t \in [s,T)$ and almost all $s \in [0,T)$, where $E$ is defined as
\begin{align}
E(t) := \int_{\Omega} \frac{\rho(\varphi(t))}{2} | \vec{v}(t) |^{2} + \frac{\eps}{2} | \nabla \varphi(t) |^{2} + \frac{d(q(t))}{\eps} \psi(\varphi(t)) + G(c(q(t))) \, dx.
\end{align}
\end{defn}
To obtain weak solutions to \eqref{Model:Josef}-\eqref{ModelJ:Bdy}, we make the following assumptions:
\begin{assump}\label{assump:Existence}
We assume that $\Omega \subset \mathbb{R}^{d}$, $d = 2,3$, is a bounded domain with $C^{2}$-boundary $\partial \Omega$.  The assumptions on the initial data $(\vec{v}_{0}, \varphi_{0}, q_{0})$ are as stated in Definition \ref{defn:weaksoln}.  Furthermore, we assume that
\begin{enumerate}
\item $\psi$, $\rho$, $\eta$, $M$, and $m$ are smooth functions, and there exist positive constants $c_{0}$, $c_{1}$, $c_{2}$, $c_{3}$, $c_{4}$, $c_{5}$, $c_{6}$, $c_{7}$ such that, for all $s,t \in \mathbb{R}$,
\begin{align*}
c_{0} < \rho(s), \eta(s) < c_{1}, \quad |\rho'(s)| + |\rho''(s)| \leq c_{2}, \quad  c_{3} \leq M(s,t), m(s) \leq c_{4}, \\
 \psi(s) \geq 0,  \quad |\psi(s)| \leq c_{5} (|s|^{3} + 1), \quad |\psi'(s)| \leq c_{5}(|s|^{2} + 1), \quad \psi(s) \geq c_{6} |s| - c_{7},
\end{align*}
with $\rho(s) = \frac{\tilde{\rho}_{+} - \tilde{\rho}_{-}}{2} s + \frac{\tilde{\rho}_{+} + \tilde{\rho}_{-}}{2}$ if $s \in [-1,1]$.  If it holds that $\rho'(\varphi)$ is not constant, then there exists a positive constant $c_{8}$ and $p \in (0,1)$ such that
\begin{align*}
|\psi'(s)| \leq c_{8} ( |s|^{p} + 1) \quad \forall s \in \mathbb{R}.
\end{align*}
\item $h$ is a smooth concave function and $d,f$ are smooth functions that satisfy the relations
\begin{align*}
d(s) = h(s) + f(s) s, \quad h'(s) = -f(s),
\end{align*}
and there exist constants $q_{\min}, q_{\max} \in \mathbb{R}$ with $q_{\min} < q_{\max}$ such that $d(s)$ is constant for $s \notin [q_{\min}, q_{\max}]$.
\item The function $c \in C^{2}(\mathbb{R})$ is strongly monotone, i.e., for some positive constant $K$,
\begin{align*}
(c(a) - c(b))(a-b) \geq K|a-b|^{2} \quad \forall a,b \in \mathbb{R}.
\end{align*} 
The composite function $\hat{G} := G \circ c \in C^{2}(\mathbb{R})$ is strictly convex and there exist positive constants $c_{9}, c_{10}$ such that
\begin{align*}
\hat{G}'(0) & = 0, \quad \hat{G}'(r) < c_{9} r, \quad \hat{G}'(t) > c_{9} t, \\
\hat{G}'(s) & = s c'(s), \quad |\hat{G}(s)| \leq c_{10} ( |s|^{2} + 1), \quad |\hat{G}'(s)| \leq c_{10}(|s| + 1),
\end{align*}
for all $s \in \mathbb{R}$, $r < 0$, $t > 0$.
\end{enumerate}
\end{assump}
We now state the existence result:
\begin{thm}[Existence of weak solutions]
Under Assumption \ref{assump:Existence}, for any $0 < T < \infty$, there exists a weak solution $(\vec{v}, \varphi, \mu, q)$ to \eqref{Model:Josef}-\eqref{ModelJ:Bdy} in the sense of Definition \ref{defn:weaksoln}.
\end{thm}
The idea of the proof is to first show the existence a weak solution $(\vec{v}^{\delta}$, $\varphi^{\delta}$, $\mu^{\delta}$, $q^{\delta})$ to a regularized version of  \eqref{Model:Josef} with an additional $\delta \partial_{t}\varphi$ on the right-hand side of \eqref{ModelJ:chem}, and an additional $\delta \Delta^{2} \vec{v}$ on the left-hand side of \eqref{ModelJ:momentum} for $\delta > 0$.  This is achieved with an semi-implicit time discretization, where the existence of time-discrete solutions $(\vec{v}^{\delta}_{N}, \varphi^{\delta}_{N}, \mu^{\delta}_{N}, q^{\delta}_{N})_{N \in \mathbb{N}}$ are established with the aid of the Leray--Schauder principle.  A crucial step is to show the compactness of $\{q^{\delta}_{N}\}_{N \in \mathbb{N}}$ in $L^{2}(Q_{T})$, which is obtained with the aid of a compactness result due to Simon \cite{Simon} and \eqref{ModelJ:surfactant}.  Then, by passing to the limit $\delta \to 0$, we obtain a weak solution to \eqref{Model:Josef}-\eqref{ModelJ:Bdy}.  For more details we refer the reader to \cite{AGW,WeberThesis}.  

\bibliographystyle{plain}
\bibliography{SPPGarcke}

\end{document}